\documentclass[fdp,a4paper,fleqn%
]{w-art}
\def\gappeq{\mathrel{\rlap {\raise.5ex\hbox{$>$}} {\lower.5ex\hbox{$\sim$}}}}
\def\lappeq{\mathrel{\rlap{\raise.5ex\hbox{$<$}} {\lower.5ex\hbox{$\sim$}}}}

\usepackage{times,cite,w-thm}
\theoremstyle{plain}

\theoremstyle{definition}

\usepackage[]{graphicx}
\begin{document}
\DOIsuffix{theDOIsuffix}
\Volume{55}
\Month{01}
\Year{2007}
\pagespan{1}{}
\Receiveddate{XXXX}
\Reviseddate{XXXX}
\Accepteddate{XXXX}
\Dateposted{XXXX}



\title[EW Symmetry Breaking]{The Electroweak Symmetry Breaking Riddle}


\author[G. Altarelli]{Guido Altarelli\inst{1}%
  \footnote{\quad E-mail:~\textsf{guido.altarelli@cern.ch}}}
            
\address[\inst{1}]{Dipartimento di Fisica `E.~Amaldi', Universit\`a di Roma Tre
and INFN, Sezione di Roma Tre, I-00146 Rome, Italy and \\ CERN, Department of Physics, Theory Unit, 
 CH--1211 Geneva 23, Switzerland}

\begin{abstract}
I present a concise review of the Higgs problem which plays a central role in particle physics today. 
The Higgs of the minimal Standard Model is so far just a conjecture that needs to be verified or discarded at the LHC. Probably the reality is more complicated. I will summarize the motivation for New Physics that should accompany or even replace the Higgs discovery and a number of its possible forms that could be revealed by the LHC.
\end{abstract}

\maketitle                   


\section{The programme of LHC physics}

Finally in November '09 the first collisions were observed at the LHC and the physics run at $3.5$ TeV per beam will soon start. The particle physics community eagerly waits for the answers that one expects from the LHC  to a number of big questions. The main physics issues at the LHC, addressed by the ATLAS and CMS collaborations, will be: 1) the experimental clarification of the Higgs sector of the electroweak (EW) theory, 2) the search for new physics at the weak scale that, on conceptual grounds, one predicts should be in the LHC discovery range, and, hopefully, 3) the identification of the particle(s) that make the Dark Matter in the Universe, in particular if those are WIMPs (Weakly Interacting Massive Particles). In addition the LHCb detector will be devoted to the study of precision B physics, with the aim of going deeper into the physics of the Cabibbo-Kobayashi-Maskawa (CKM) matrix and of CP violation. The LHC will also devote a number of runs to accelerate heavy ions and the ALICE collaboration will study their collisions for an experimental exploration of the QCD phase diagram.

\section{The Higgs Problem}

The experimental verification of the Standard Model (SM) \cite{sprin} cannot be considered complete until the predicted physics of the  Higgs sector is not established by experiment \cite{djou1}. Indeed the Higgs problem is really central in particle physics today \cite{wells}. In fact, the Higgs sector is directly related to most of the major open problems of particle physics, like the flavour problem and the hierarchy problem, the latter strongly suggesting the need for new physics near the weak scale, which could also clarify the Dark Matter identity. 

It is clear that the fact that some sort of Higgs mechanism is at work has already been established. The longitudinal degrees of freedom for the W and the Z are borrowed from the Higgs sector and are an evidence for it. In fact the couplings of quarks and leptons to
the weak gauge bosons W$^{\pm}$ and Z are indeed experimentally found to be precisely those
prescribed by the gauge symmetry.  To a lesser
accuracy the triple gauge vertices $\gamma$WW and ZWW have also
been found in agreement with the specific predictions of the
$SU(2)\bigotimes U(1)$ gauge theory. This means that it has been
verified that the gauge symmetry is unbroken in the vertices of the
theory: all currents and charges are indeed symmetric. Yet there is obvious
evidence that the symmetry is instead badly broken in the
masses. The W or the Z with longitudinal polarization that are observed are not present in an unbroken gauge theory (massless spin-1 particles, like the photon, are transversely polarized). Not only the W and the Z have large masses, but the large splitting of, for example,  the t-b doublet shows that even the global weak SU(2) is not at all respected by the fermion spectrum. Symmetric couplings and totally non symmetric spectrum is a clear signal of spontaneous
symmetry breaking and its implementation in a gauge theory is via the Higgs mechanism. The big remaining questions are about the nature and the properties of the Higgs particle(s). 

The LHC has been designed to solve the Higgs problem. A strong argument indicating that the solution of the Higgs problem cannot be too far away is the fact that, in the absence of a Higgs particle or of an alternative mechanism, violations of unitarity appear in scattering amplitudes involving longitudinal gauge bosons (those most directly related to the Higgs sector) at energies in the few TeV range \cite{ref:unit}. A crucial question for the LHC is to identify the mechanism that avoids the unitarity violation: is it one or more Higgs bosons or some new vector boson (like additional gauge bosons WÕ, ZÕ or Kaluza-Klein recurrences or resonances from a strong sector)?

It is well known that in the SM with only one Higgs doublet a lower limit on
$m_H$ can be derived from the requirement of vacuum stability (i.e. that the quartic Higgs coupling $\lambda$ does not turn negative in its running up to a large scale $\Lambda$) or, in milder form, of a moderate instability, compatible with the lifetime of the Universe  \cite{ref:isid}. The Higgs mass enters because it fixes the initial value of the quartic Higgs coupling $\lambda$. For the experimental value of $m_t$ the lower limit is below the direct experimental bound for $\Lambda \sim $ a few TeV and is $M_H> 130$ GeV for $\Lambda \sim M_{Pl}$. Similarly an upper bound on $m_H$ (with mild dependence
on $m_t$) is obtained, as described in \cite{ref:hri}, from the requirement that for $\lambda$ no Landau pole appears up to the scale $\Lambda$, or in simpler terms, that the perturbative description of the theory remains valid up to  $\Lambda$. The upper limit on the Higgs mass in the SM is clearly important for assessing the chances of success of the LHC as an accelerator designed to solve the Higgs problem. Even if $\Lambda$ is as small as ~a few TeV the limit is $m_H < 600-800~$GeV and becomes $m_H < 180~$GeV for $\Lambda \sim M_{Pl}$. 

In conclusion it looks very likely that the LHC can clarify the problem of the electroweak symmetry breaking mechanism. It has been designed for it!

\section{Precision Tests of the Standard Electroweak Theory}

The most precise tests of the electroweak theory apply to the QED sector. The anomalous magnetic moments of the electron and of the muon are among the most precise measurements in the whole of physics. Recently there have been new precise measurements of $a$ for the electron \cite{ref:ae1} and the muon \cite{ref:amu} ($a = (g-2)/2$). The QED part has been computed analytically for $i=1,2,3$, while for $i=4$ there is a numerical calculation (see, for example, \cite{ref:kino}). Some terms for $i=5$ have also been estimated for the muon case. The weak contribution is from $W$ or $Z$ exchange. The hadronic contribution is from vacuum polarization insertions and from light by light scattering diagrams.  For the electron case the weak contribution is essentially negligible and the hadronic term does not introduce an important uncertainty.  As a result the $a_e$ measurement can be used to obtain the most precise determination of the fine structure constant \cite{ref:ae2}. In the muon case the experimental precision is less by about 3 orders of magnitude, but the sensitivity to new physics effects is typically increased by a factor $(m_\mu/m_e)^2 \sim 4^.10^4$. The dominant theoretical ambiguities arise from the hadronic terms in vacuum polarization and in light by light scattering. If the vacuum polarization terms are evaluated from the $e^+e^-$ data a discrepancy of $\sim 3 \sigma$ is obtained (the $\tau$ data would indicate better agreement, but the connection to $a_\mu$ is less direct and recent new data have added solidity to the $e^+e^-$ route)\cite{ref:amu2}. Finally, we note that, given the great accuracy of the $a_\mu$ measurement and the estimated size of the new physics contributions, for example from SUSY, it is not unreasonable that a first signal of new physics would appear in this quantity.

The results of the electroweak precision tests also including the measurements of $m_t$, $m_W$ and the searches for new physics at the Tevatron form a very stringent set of precise constraints \cite{ref:ewg} to compare with the Standard Model (SM) or with
any of its conceivable extensions. When confronted with these results, on the whole the SM performs rather
well, so that it is fair to say that no clear indication for new physics emerges from the data \cite{ref:AG}.  But the
Higgs sector of the SM is still very much untested. What has been
tested is the relation $M_W^2=M_Z^2\cos^2{\theta_W}$, modified by small, computable
radiative corrections. This relation means that the effective Higgs
(be it fundamental or composite) is indeed a weak isospin doublet.
The Higgs particle has not been found but in the SM its mass can well
be larger than the present direct lower limit $m_H > 114.4$~GeV
obtained from direct searches at LEP-2.  The radiative corrections
computed in the SM when compared to the data on precision electroweak
tests lead to a clear indication for a light Higgs, not too far from
the present lower bound. The exact upper limit for $m_H$ in the SM depends on the value of the top quark mass $m_t$ (the one-loop radiative corrections are quadratic in $m_t$ and logarithmic in $m_H$). The measured value of $m_t$ went down recently (as well as the associated error) according to the results of Run II at the Tevatron. The CDF and D0 combined value is at present $m_t~= 173.1~\pm~1.3~GeV$. As a consequence the present limit on $m_H$ is quite stringent: $m_H < 186~GeV$ (at $95\%$ c.l., after including the information from the 114.4 GeV direct bound)  \cite{ref:ewg}.  

In the Higgs search the Tevatron is now reaching the SM sensitivity. The most recent limit, reported near the end of 2009, is somewhat weaker: $163 < m_H < 166$ GeV \cite{SJindariani}. The goal at Fermilab is to collect $12~fb^{-1}$ of luminosity by 2011 and possibly exclude $115 < m_H < 185$ GeV.

\section{The Physics of Flavour}

Another domain where the SM is really in good agreement with the data is flavour physics (actually too good in comparison with the general expectation before the experiments). In the last decade great progress in different areas of flavour physics has been achieved. In the quark sector, the amazing results of a generation of frontier experiments, performed at B factories and at accelerators, have become available. 
The hope of the B-decay experiments was to detect departures from the CKM picture of mixing and of CP violation as  signals of new physics. At present the available results on B mixing and CP violation on the whole agree very well with the SM predictions based on the CKM matrix \cite{rept}. A few interesting ÒtensionsÓ at the 2-3 $\sigma$ level should be monitored closely in the future: $\sin{2\beta}$ from $B_d \rightarrow J/\Psi K^0$ versus $\epsilon_K$ and $V_{ub}$ (which, however, in my opinion, is probably due to an underestimate of theoretical errors, particularly on the determination of $V_{ub}$), $\beta_s$ measured by CDF, $D_0$
in $B_s \rightarrow J/\Psi \phi$ and $B\rightarrow \tau \nu$. But certainly the amazing performance of the SM in flavour changing  and/or CP violating transitions in K and B decays poses very strong constraints on all proposed models of new physics \cite{isid}. For example, if one adds to the SM effective non renormalizable operators suppressed by powers of a scale $\Lambda$ one generally finds that experiments indicate very large values of $\Lambda$, much above the few TeV range indicated by the hierarchy problem. Only if one assumes that the relevant new physics effects are suppressed at the tree level and mainly occur at the loop level and that, in addition, the new physics inherits the same SM protections against flavour changing neutral currents (like the GIM mechanism and small $V_{CKM}$ factors) as, for example, in Minimal Flavour Violation models \cite{isid},  that one obtains bounds on  $\Lambda$ in the few TeV range.

In the leptonic sector the study of neutrino oscillations has led to the discovery that at least two neutrinos are not massless and to the determination of the mixing matrix \cite{revnu}. Neutrinos are not all massless but their masses are very small (at most a fraction of $eV$). The neutrino spectrum could be either of the normal hierarchy type (with the solar doublet below), or of the inverse hierarchy type (with the solar doublet above). Probably masses are small because $\nu$Õs are Majorana fermions, and, by the see-saw mechanism, their masses are inversely proportional to the large scale $M$ where lepton number ($L$) non conservation occurs (as expected in GUT's). Indeed the value of $M\sim m_{\nu R}$ from experiment is compatible with being close to $M_{GUT} \sim 10^{14}-10^{15}GeV$, so that neutrino masses fit well in the GUT picture and actually support it. The interpretation of neutrinos as Majorana particles enhances the importance of experiments aimed at the detection of neutrinoless double beta decay and a huge effort in this direction is underway.  It was realized that decays of heavy $\nu_R$ with CP and L non conservation can produce a B-L asymmetry (which is unchanged by instanton effects at the electroweak scale). The range of neutrino masses indicated by neutrino phenomenology turns out to be perfectly compatible with the idea of baryogenesis via leptogenesis \cite{ref:buch}. This elegant model for baryogenesis has by now replaced the idea of baryogenesis near the weak scale, which has been strongly disfavoured by LEP. It is remarkable that we now know the neutrino mixing matrix with good accuracy \cite{datanu}. Two mixing angles are large and one is small. The atmospheric angle $\theta_{23}$ is large, actually compatible with maximal but not necessarily so. The solar angle $\theta_{12}$ (the best measured) is large, $\sin^2{\theta_{12}}\sim 0.3$, but certainly not maximal. The third angle $\theta_{13}$, strongly limited mainly by the CHOOZ experiment, has at present a $3\sigma$ upper limit given by about $\sin^2{\theta_{13}}\leq 0.04$. It is a fact that, to a precision comparable with the measurement accuracy, the Tri-Bimaximal (TB) mixing pattern ($\sin^2{\theta_{12}}\sim 1/3$, $\sin^2{\theta_{23}}\sim 1/2$ and $\sin^2{\theta_{13}} = 0$) \cite{Harr} is well approximated by the data. If this experimental result is not a mere accident but a real indication that a dynamical mechanism is at work to guarantee the validity of TB mixing in the leading approximation, corrected by small non leading terms, then non abelian discrete flavor groups emerge as the main road to an understanding of this mixing pattern \cite{rmp}. Indeed the entries of the TB mixing matrix are clearly suggestive of "rotations" by simple, very specific angles. In fact the group $A_4$, the simplest group used to explain TB mixing, is defined as the group of rotations that leave a regular rigid tetrahedron invariant. The non conservation of the three separate lepton numbers and the large leptonic mixing angles make it possible that processes like $\mu \rightarrow e \gamma$ or $\tau \rightarrow \mu \gamma$ might be observable, not in the SM but in extensions of it like the MSSM. Thus, for example, the outcome of the now running experiment MEG at PSI \cite{MEG} aiming at improving the limit on $\mu \rightarrow e \gamma$ by 1 or 2 orders of magnitude, is of great interest. 

\section{Outlook on Avenues beyond the Standard Model}

No signal of new physics has been
found neither in electroweak precision tests nor in flavour physics. Given the success of the SM why are we not satisfied with that theory? Why not just find the Higgs particle,
for completeness, and declare that particle physics is closed? The reason is that there are
both conceptual problems and phenomenological indications for physics beyond the SM. On the conceptual side the most
obvious problems are the proliferation of parameters, the puzzles of family replication and of flavour hierarchies, the fact that quantum gravity is not included in the SM and the related hierarchy problem. Among the main
phenomenological hints for new physics we can list Dark Matter, the quest for Grand Unification and coupling constant merging, neutrino masses (explained in terms of L non conservation), 
baryogenesis and the cosmological vacuum energy (a gigantic naturalness problem).

We know by now  that  the  Universe is flat and most of it is not made up of known forms of matter \cite{oli}: while $\Omega_{tot} \sim 1$ and $\Omega_{matter} \sim 0.3$, the normal baryonic matter is only $\Omega_{baryonic} \sim 0.044$, where $\Omega$ is the ratio of the density to the critical density. Most of the energy in the Universe is Dark Matter (DM) and Dark Energy (DE) with $\Omega_{\Lambda} \sim 0.7$. We also know that most of DM must be cold (non relativistic at freeze-out) and that significant fractions of hot DM are excluded. Neutrinos are hot DM (because they are ultrarelativistic at freeze-out) and indeed are not much cosmo-relevant: $\Omega_{\nu} \lappeq 0.015$. The identification of DM is a task of enormous importance for both particle physics and cosmology. The LHC has good chances to solve this problem in that it is sensitive to a large variety of WIMP's (Weekly Interacting Massive Particles). It is a fact that WIMP's with masses in the 10 GeV-1 TeV range and typical EW cross-sections turn out to contribute terms of $o(1)$ to $\Omega$. This is a formidable hint in favour of WIMP's as DM candidates. By comparison, axions are also DM candidates but their mass and couplings must be tuned for this purpose. If really some sort of WIMP's are a main component of DM they could be discovered at the LHC and this will be a great service of particle physics to cosmology. Active searches in non-accelerator experiments are under way. Some hints of possible signals have been reported: e.g. annual modulations (DAMA/LIBRA at Gran Sasso \cite{ber}), $e^+$ and/or $e^+e^-$ excess in cosmic ray detectors, e.g. in PAMELA \cite{pam} and ATIC \cite{atic} (but the ATIC excess has not been confirmed by Fermi-LAT \cite{fer}). If those effects are really signals for DM they would indicate particularly exotic forms of DM \cite{exdm}. But for the PAMELA effect an astrophysical explanation in terms of relatively close pulsars appears as a plausible alternative \cite{astr}.

The computed evolution with energy
of the effective gauge couplings clearly points towards the unification of the electro-weak and strong forces (Grand Unified Theories: GUT's) at scales of energy $M_{GUT}\sim  10^{15}-10^{16}~ GeV$ which are close to the scale of quantum gravity, $M_{Pl}\sim 10^{19}~ GeV$.  One is led to imagine  a unified theory of all interactions also including gravity (at present superstrings provide the best attempt at such a theory). Thus GUT's and the realm of quantum gravity set a very distant energy horizon that modern particle theory cannot ignore. Can the SM without new physics be valid up to such large energies? One can imagine that some of the obvious problems of the SM could be postponed to the more fundamental theory at the Planck mass. For example, the explanation of the three generations of fermions and the understanding of the pattern of fermion masses and mixing angles can be postponed. But other problems must find their solution in the low energy theory. In particular, the structure of the SM could not naturally explain the relative smallness of the weak scale of mass, set by the Higgs mechanism at $\mu\sim 1/\sqrt{G_F}\sim  250~ GeV$  with $G_F$ being the Fermi coupling constant. This so-called hierarchy problem \cite{hie} is due to the instability of the SM with respect to quantum corrections. This is related to the presence of fundamental scalar fields in the theory with quadratic mass divergences and no protective extra symmetry at $\mu=0$. For fermion masses, first, the divergences are logarithmic and, second, they are forbidden by the $SU(2)\bigotimes U(1)$ gauge symmetry plus the fact that at $m=0$ an additional symmetry, i.e. chiral  symmetry, is restored. Here, when talking of divergences, we are not worried of actual infinities. The theory is renormalizable and finite once the dependence on the cut-off $\Lambda$ is absorbed in a redefinition of masses and couplings. Rather the hierarchy problem is one of naturalness \cite{giu}. We can look at the
cut-off as a parameterization of our ignorance on the new physics that will modify the theory at large energy
scales. Then it is relevant to look at the dependence of physical quantities on the cut-off and to demand that no unexplained enormously accurate cancellations arise. 
In fact, the hierarchy problem can be put in quantitative terms: loop corrections to the higgs mass squared are quadratic in the cut-off $\Lambda$. The most pressing problem is from the top loop.
 With $m_h^2=m^2_{bare}+\delta m_h^2$ the top loop gives 
 \begin{eqnarray}
\delta m_{h|top}^2\sim -\frac{3G_F}{2\sqrt{2} \pi^2} m_t^2 \Lambda^2\sim -(0.2\Lambda)^2 \label{top}
\end{eqnarray}
If we demand that the correction does not exceed the light Higgs mass indicated by the precision tests, $\Lambda$ must be
close, $\Lambda\sim o(1~TeV)$. Similar constraints also arise from the quadratic $\Lambda$ dependence of loops with gauge bosons and
scalars, which, however, lead to less pressing bounds. So the hierarchy problem demands new physics to be very close (in
particular the mechanism that quenches the top loop). Actually, this new physics must be rather special, because it must be
very close, yet its effects are not clearly visible in precision electroweak tests - the "LEP Paradox" \cite{ref:BS} - now also accompanied by a similar "flavour paradox" \cite{isid}. Examples of proposed classes of solutions for the hierarchy problem are described in the following sections  \cite{CGrojean}.

\section{Supersymmetry: the standard way beyond the SM}

Models based on supersymmetry (SUSY)\cite{Martin} are the most developed and widely known. In the limit of exact boson-fermion symmetry the quadratic divergences of bosons cancel, so that
only logarithmic divergences remain. However, exact SUSY is clearly unrealistic. For approximate SUSY (with soft breaking terms),
which is the basis for all practical models, $\Lambda$ in eq.(\ref{top}) is essentially replaced by the splitting of SUSY multiplets. In particular, the top loop is quenched by partial cancellation with s-top exchange, so the s-top cannot be too heavy. 

The Minimal SUSY Model (MSSM) is the extension of the SM with minimal particle content. To each ordinary particle a s-particle is associated with 1/2 spin difference: to each helicity state of a spin 1/2 fermion of the SM a scalar is associated (for example, the electron states $e_L$ and $e_R$ correspond to 2 scalar s-electron states). Similarly to each ordinary boson a s-fermion is associated: for example to each gluon a gluino (a Majorana spin 1/2 fermion) is related. Why not even a single s-particle was seen so far? A clue: observed particles are those whose mass is forbidden by $SU(2) \bigotimes U(1)$. When SUSY is broken but $SU(2) \bigotimes U(1)$ is unbroken s-particles get a mass but particles remain massless. Thus if SUSY breaking is large we understand that no s-particles have been observed yet. It is an important fact that two Higgs doublets, $H_u$ and $H_d$, are needed in the MSSM with their corresponding spin 1/2 s-partners, to give mass to the up-type and to the down-type fermions, respectively. This duplication is needed for cancellation of the chiral anomaly and also because the SUSY rules forbid that $H_d=H^\dagger_u$ as is the case in the the SM. The ratio of their two vacuum expectation values $\tan{\beta}=v_u/v_d$ (with the SM VEV $v$ being given by $v=\sqrt{v_u^2+v_d^2}$) plays an important role for phenomenology.

The most general MSSM symmetric renormalizable lagrangian would contain terms that violate baryon $B$ and lepton $L$ number conservation (which in the SM, without $\nu_R$, are preserved at the renormalizable level, so that they are "accidental" symmetries). To eliminate those terms it is sufficient to invoke a discrete parity, $R$-parity, whose origin is assumed to be at a more fundamental level, which is $+1$ for ordinary particles and $-1$ for s-partners.
The consequences of $R$-parity are that s-particles are produced in pairs at colliders, 
the lightest s-particle is absolutely stable (it is called the Lightest SUSY Particle, LSP, and is a good candidate for Dark Matter) and s-particles decay into a final state with an odd number of s-particles (and, ultimately, in the decay chain there will be the LSP).

The necessary SUSY breaking, whose origin is not clear, can be phenomenologically introduced through soft terms (i. e. with operator dimension $< 4$) that do not spoil the good convergence properties of the theory (renormalizability and non renormalization theorems 
are maintained). We denote by $m_{soft}$ the mass scale of the soft SUSY breaking terms. The most general soft terms compatible with the SM gauge symmetry and with $R$-parity conservation introduce more than one hundred new parameters. In general new sources of flavour changing neutral currents (FCNC) and of CP violation are introduced e.g. from s-quark mass matrices. Universality (proportionality of the mass matrix to the identity matrix for each charge sector) and/or alignment (near diagonal mass matrices) must be assumed at a large scale, but renormalization group running can still produce large effects. The MSSM does provide a viable flavour framework in the assumption of $R$-parity conservation, universality of soft masses and proportionality of trilinear soft terms to the SM Yukawas (still broken by renormalization group running). As already mentioned, observable effects in the lepton sector are still possible (e.g. $\mu \rightarrow e \gamma$ or $\tau \rightarrow \mu \gamma$) \cite{mas}. This is made even more plausible by large neutrino mixings.

How can SUSY breaking be generated? Conventional spontaneous symmetry breaking of SUSY cannot occur within the MSSM and also
in simple extensions of it. Probably the soft terms of the MSSM arise indirectly or radiatively 
(loops) rather than from tree level renormalizable couplings. The prevailing idea is that it happens in a "hidden sector" through non renormalizable interactions and is communicated to the visible sector by some interactions. Gravity is a plausible candidate for the hidden sector. Many theorists consider SUSY as established at the Planck
scale $M_{Pl}$. So why not to use it also at low energy to fix the hierarchy problem, if at all possible? It is interesting that viable models exist. Suitable soft terms indeed arise from supergravity when it is spontaneoulsly broken. Supergravity is a non renormalizable SUSY theory of quantum gravity\cite{Martin}. The SUSY partner of the spin-2 graviton $g_{\mu \nu}$ is the spin-3/2 gravitino $\Psi_{i \mu}$ (i: spinor index, $\mu$: Lorentz index). The gravitino is the gauge field associated to the SUSY generator. When SUSY is broken the gravitino takes mass by absorbing the 2 goldstino components (super-Higgs mechanism). In gravity mediated SUSY breaking  typically the gravitino mass $m_{3/2}$ is of order $m_{soft}$ (the scale of mass of the soft breaking terms) and, on dimensional ground, both are given by $m_{3/2}\sim m_{soft}\sim \langle F \rangle/M_{Pl}$, where $F$ is the dimension 2 auxiliary field that takes a vacuum expectation value $\langle F \rangle$ in the hidden sector (the denominator $M_{Pl}$ arises from the gravitational coupling that transmits the breaking down to the visible sector). For $m_{soft}\sim 1~{\rm TeV}$, the scale of SUSY breaking is very large of order $\sqrt{\langle F \rangle}\sim \sqrt{m_{soft}M_{Pl}}\sim 10^{11}~ {\rm GeV}$. With ~TeV mass and gravitational coupling the gravitino is not relevant for LHC physics but perhaps for  cosmology (it could be the LSP and a Dark Matter candidate). In gravity mediation the neutralino is the typical LSP and an excellent Dark Matter candidate. A lot of missing energy is a signature for gravity mediation. 

\begin{figure}
\centerline{\includegraphics[height=4in]{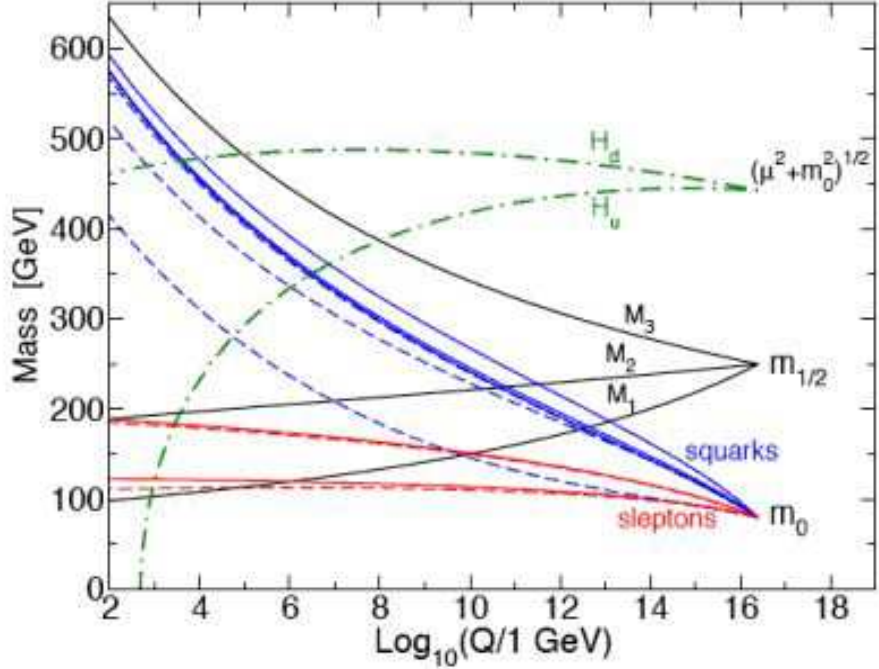}}     
\caption{A SUSY spectrum generated by universal boundary conditions at the GUT scale \label{sugra}}
\end{figure}

Different mechanisms of SUSY breaking are also being considered. In one alternative scenario\cite{gau} the (not so
much) hidden sector is connected to the visible one by ÒmessengerÓ heavy fields, with mass $M_{mess}$, which share ordinary gauge interactions and thus, in amplitudes involving only external light particles, appear in 
loops  so that $m_{soft}\sim \frac{\alpha_i}{4\pi}\frac{\langle F \rangle}{M_{mess}}$. Both gaugino and s-fermion masses are of order $m_{soft}$. Messengers can be taken in complete SU(5) representations, like 5+$\bar 5$, so that coupling unification is not spoiled. As gauge interactions are much
stronger than gravitational interactions, the SUSY breaking scale can be much smaller, as low as $\sqrt{\langle F \rangle}\sim M_{mess}\sim 10-100~{\rm TeV}$. It follows that the gravitino is very light (with mass of order or below $1~{\rm eV}$ typically) and, in these models, always is the LSP. Its couplings are observably large because the gravitino couples to SUSY particle multiplets through its spin 1/2 goldstino components. Any SUSY particle will eventually decay into the gravitino. But the decay of the next-to-the lightest SUSY particle (NLSP) could be extremely slow, with a travel path at the LHC from microscopic to astronomical distances. The main appeal of gauge mediated models is a better protection against FCNC: if one starts at $M_{mess}$ with sufficient universality/alignment then the very limited  interval for renormalization group running down to the EW scale does not spoil it. Indeed at $M_{mess}$ there is approximate alignment because the mixing
parameters $A_{u.d,l}$ in the soft breaking lagrangian are of dimension of mass and arise at two loops, so that they are suppressed.

What is unique to SUSY with respect to most other extensions of the SM is that SUSY models are well defined and computable up to $M_{Pl}$ and, moreover, are not only compatible but actually 
quantitatively supported by coupling unification and GUT's. At present the most direct
phenomenological evidence in favour of SUSY is obtained from the unification of couplings in GUT's.
Precise LEP data on $\alpha_s(m_Z)$ and $\sin^2{\theta_W}$ show that
standard one-scale GUT's fail in predicting $\alpha_s(m_Z)$ given $\sin^2{\theta_W}$ 
and $\alpha(m_Z)$ while SUSY GUT's are compatible with the present, very precise,
experimental results (of course, the ambiguities in the MSSM prediction are larger than for the SM case because of our ignorance of the SUSY spectrum). If one starts from the known values of
$\sin^2{\theta_W}$ and $\alpha(m_Z)$, one finds\cite{LP} for $\alpha_s(m_Z)$ the results:
$\alpha_s(m_Z) = 0.073\pm 0.002$ for Standard GUT's and $\alpha_s(m_Z) = 0.129\pm0.010$ for SUSY~ GUT's
to be compared with the world average experimental value $\alpha_s(m_Z) =0.118\pm0.002$\cite{pdg}. Another great asset of SUSY GUT's
is that proton decay is much slowed down with respect to the non SUSY case. First, the unification mass $M_{GUT}\sim~\rm{few}~
10^{16}~GeV$, in typical SUSY GUT's, is about 20 times larger than for ordinary GUT's. This makes p decay via gauge
boson exchange negligible and the main decay amplitude arises from dim-5 operators with higgsino exchange, leading to a
rate close but still compatible with existing bounds (see, for example,\cite{AFM}).

\begin{figure}
\centerline{\includegraphics[height=4in]{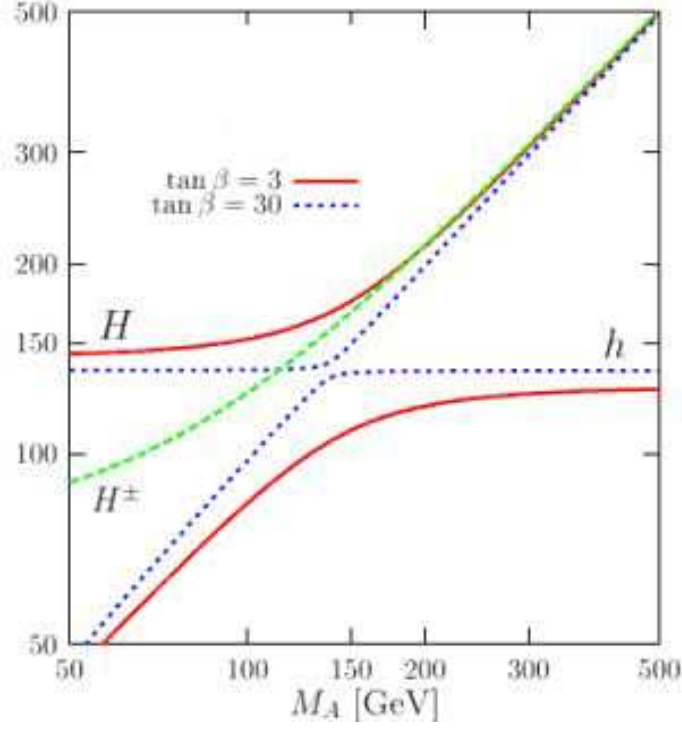}}     
\caption{The MSSM Higgs spectrum as function of $m_A$: $h$ is  the lightest Higgs, $H$ and $A$ are the heavier neutral scalar and pseudoscalar Higgs, respectively,  and $H^\pm$ are the charged Higgs bosons. The curves refer to $m_t=178~{\rm GeV}$ and large top mixing $A_t$ \label{higgs}}
\end{figure}

By imposing on the MSSM model universality constraints at $M_{GUT}$  one obtains a drastic reduction in the number of parameters at the price of more rigidity and model dependence (see Figure~\ref{sugra}\cite{Martin}). This is the SUGRA or CMSSM (C for "constrained") limit \cite{Martin}.
An interesting exercise is to repeat the fit of precision tests in the CMSSM, also including the additional data on the muon $(g-2)$, the Dark Matter relic density and the $b\rightarrow s \gamma$ rate. The result \cite{sus} is that the central value of the lightest Higgs mass $m_h$ goes up (in better harmony with the bound from direct searches) with moderately large $tan\beta$ and relatively light SUSY spectrum.

\begin{figure}
\centerline{\includegraphics[height=4in]{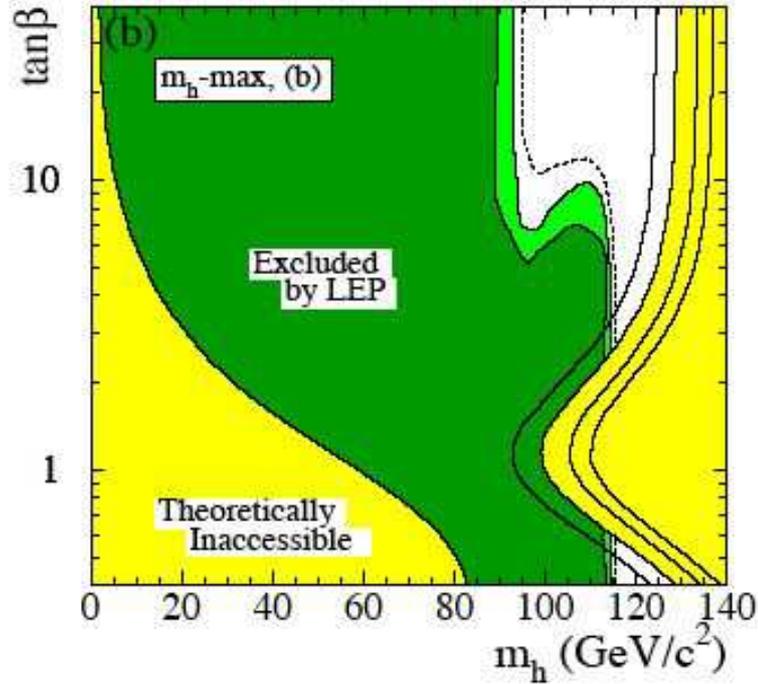}}     
\caption{Experimental limits in the $\tan{\beta}-m_h$ plane from LEP. With h one denotes the lightest MSSM Higgs boson. \label{limits}}
\end{figure}

In spite of all these virtues it is true that the lack of SUSY signals at LEP and the lower limit on $m_H$ pose problems
for the MSSM. The predicted spectrum of Higgs particles in the MSSM is shown in Figure~\ref{higgs}\cite{djou}. As apparent from the figure the lightest Higgs particle is predicted in the MSSM to be below $m_h\lappeq~130~GeV$ (with the esperimental value of $m_t$ going down the upper limit is slightly decreased). In fact, at tree level $m^2_h=m^2_Z\cos^2{2\beta}$ and it is only through radiative corrections that $m_h$ can increase beyond $m_Z$:
\begin{eqnarray}
\Delta m_h^2\sim \frac{3G_F}{\sqrt{2} \pi^2} m_t^4\log{\frac{m_{\tilde{t_1}}m_{\tilde{t_2}}}{m^2_t}} \label{stop}
\end{eqnarray}
Here $\tilde{t}_{1,2}$ are the s-top mass eigenstates. 
The direct limit on $m_h$ from the Higgs search at LEP, shown in Figure~\ref{limits} \cite{lephwg}, considerably restricts the available parameter space of the MSSM requiring relatively large $\tan\beta$
and heavy s-top quarks. Stringent naturality constraints also follow from imposing that the EW breaking occurs at the right energy scale. A great feature of SUSY models is that the EW breaking is induced by the running of the $H_u$ mass
starting from a common scalar mass $m_0$ at $M_{GUT}$ (see Figure~\ref{sugra}). The squared Z mass $m_Z^2$ can be expressed as a linear
combination of the SUSY parameters $m_0^2$, $m_{1/2}^2$, $A^2_t$, $\mu^2$,... with known coefficients. Barring
cancellations that need fine tuning, the SUSY parameters, hence the SUSY s-partners, cannot be too heavy. The LEP limits,
in particular the chargino lower bound $m_{\chi+}\gappeq~100~GeV$, are sufficient to eliminate an important region of the
parameter space, depending on the amount of allowed fine tuning. For example, models based on gaugino universality at the
GUT scale, like the CMSSM, need a fine tuning by at least a factor of ~20  \cite{ross}. Without gaugino
universality \cite{kane} the strongest limit remains on the gluino mass: the relation reads $m_Z^2\sim 0.7~m_{gluino}^2+\dots$ and is still
compatible with the present limit $m_{gluino} \gappeq~250-300~GeV$ from the Tevatron.

So far we have discussed the case of the MSSM with minimal particle content. Of course, minimality is only a simplicity assumption that could possibly be relaxed \cite{extsusy}. For example, adding an additional Higgs singlet S considerably helps in addressing naturalness constraints (Next-to Minimal SUSY SM: NMSSM) \cite{nmssm},\cite{barbie}. An additional singlet can also help solving the "$\mu$- problem"\cite{Martin} . In the exact SUSY and gauge symmetric limit there is a single parameter with dimension of mass in the superpotential. The $\mu$ term in the superpotential is of the form $W_{\mu term}=\mu H_uH_d$. The mass $\mu$, which contributes to the Higgs sector masses, must be of order $m_{soft}$ for phenomenological reasons. The problem is to justify this coincidence, because $\mu$ could in principle be much larger given that it already appears at the symmetric level. A possibility is to forbid the $\mu$ term by a suitable symmetry in the SUSY unbroken limit and then generate it together with the SUSY breaking terms. For example, one can introduce a discrete parity that 
forbids the $\mu$ term. Then Giudice and Masiero \cite{gium} have observed that in general, the low energy limit of supergravity, also induces a SUSY conserving $\mu$ term together with the soft SUSY breaking terms and of the same order. A different phenomenologically appealing possibility is to replace $\mu$ with the VEV of a new singlet scalar field S, thus enlarging the Higgs sector as in the NMSSM.

In conclusion the main SUSY virtues are that the hierarchy problem is drastically reduced, the model agrees with the EW data,  is consistent and computable up to $M_{Pl}$, is well compatible and indeed supported by GUT's, has good Dark Matter candidates and, last not least, is testable at the LHC. The delicate points for SUSY are the origin of SUSY breaking and of R-parity, the $\mu$-problem, the flavour problem and the need of sizable fine tuning. 

\section{Technicolor}

The Higgs system is a condensate of new fermions. There are no fundamental scalar Higgs sector, hence no
quadratic devergences associated to the $\mu^2$ mass in the scalar potential. This mechanism needs a very strong binding force,
$\Lambda_{TC}\sim 10^3~\Lambda_{QCD}$. It is  difficult to arrange that such nearby strong force is not showing up in
precision tests. Hence this class of models has been disfavoured by LEP, although some special class of models have been devised a posteriori, like walking TC, top-color assisted TC etc \cite{ref:L-C}. 

\section{Little Higgs models}

The failure to discover SUSY at LEP has given further impulse to the quest for new ideas on physics beyond the SM. In "little Higgs" models \cite{schm} the symmetry of the SM is extended to a suitable global group G that also contains some
gauge enlargement of $SU(2)\bigotimes U(1)$, for example $G\supset [SU(2)\bigotimes U(1)]^2\supset SU(2)\bigotimes
U(1)$. The Higgs particle is a pseudo-Goldstone boson of G that only takes mass at 2-loop level, because two distinct
symmetries must be simultaneously broken for it to take mass,  which requires the action of two different couplings in
the same diagram. Then in the relation eq.(\ref{top})
between
$\delta m_h^2$ and
$\Lambda^2$ there is an additional coupling and an additional loop factor that allow for a bigger separation between the Higgs
mass and the cut-off. Typically, in these models one has one or more Higgs doublets at $m_h\sim~0.2~{\rm TeV}$, and a cut-off at
$\Lambda\sim~10~{\rm TeV}$. The top loop quadratic cut-off dependence is partially canceled, in a natural way guaranteed by the
symmetries of the model, by a new coloured, charge 2/3, vectorlike quark $\chi$ of mass around $1~{\rm TeV}$ (a fermion not a scalar
like the s-top of SUSY models). Certainly these models involve a remarkable level of group theoretic virtuosity. However, in
the simplest versions one is faced with problems with precision tests of the SM \cite{prob}. These problems can be fixed by complicating the model \cite{Ch}: one can introduce a parity symmetry, T-parity, and additional "mirror" fermions.  T-parity interchanges the two $SU(2)\bigotimes
U(1)$ groups: standard gauge bosons are T-even while heavy ones are T-odd. As a consequence no tree level contributions from heavy $WÕ$ and $ZÕ$ appear in processes with external SM particles. 
Therefore all corrections to EW observables only arise at loop level. A good feature of T-parity is that, like for R-parity in the MSSM, the lightest T-odd particle is stable (usually a B') and can be a candidate for Dark Matter (missing energy would here too be a signal) and T-odd particles are produced in pairs (unless T-parity is not broken by anomalies \cite{hill}). Thus the model could work but, in my opinion, the real limit of
this approach is that it only offers a postponement of the main problem by a few TeV, paid by a complete loss of
predictivity at higher energies. In particular all connections to GUT's are lost. Still it is very useful to offer to experiment a different example of possible new physics and signals to look for \cite{sign}.

\section{Extra dimensions}

Extra dimension  models are  among the most interesting new directions in model building. Early formulations were based on "large" extra dimensions \cite{led},\cite{Jo}. These are models with factorized metric: $ds^2=\eta_{\mu
\nu}dx^{\mu}dx^{\nu}+h_{ij}(y)dy^idy^j$, where $y^{i,j}$ denote the extra dimension coordinates and indices. Large
extra dimension models propose to solve the hierarchy problem by bringing gravity  from $M_{Pl}$ down to $m\sim~o(1~{\rm TeV})$ where
$m$ is the string scale. Inspired by string theory one assumes that some compactified extra dimensions are sufficiently large
and that the SM fields are confined to a 4-dimensional brane immersed in a d-dimensional bulk while gravity, which feels the
whole geometry, propagates in the bulk. We know that the Planck mass is large just because gravity is weak: in fact $G_N\sim
1/M_{Pl}^2$, where
$G_N$ is Newton constant. The new idea is that gravity appears so weak because a lot of lines of force escape in extra
dimensions. Assume you have $n=d-4$ extra dimensions with compactification radius $R$. For large distances, $r>>R$, the
ordinary Newton law applies for gravity: in natural units, the force between two units of mass is $F\sim G_N/r^2\sim 1/(M_{Pl}^2r^2)$. At short distances,
$r\lappeq R$, the flow of lines of force in extra dimensions modifies Gauss law and $F^{-1}\sim m^2(mr)^{d-4}r^2$. By
matching the two formulas at $r=R$ one obtains $(M_{Pl}/m)^2=(Rm)^{d-4}$. For $m\sim~1~TeV$ and $n=d-4$ one finds that
$n=1$ is excluded ($R\sim 10^{15} {\rm cm}$), for $n=2~R$  is very marginal and also at the edge of present bounds $R\sim~1~ {\rm mm}$ on departures from Newton law\cite{cav}, while for $n=4,6$,
$R\sim~10^{-9}, 10^{-12}~{\rm cm}$ and these cases are not excluded.   

A generic feature of extra dimensional models is the occurrence of Kaluza-Klein (KK) modes.
Compactified dimensions with periodic boundary conditions, like the case of quantization in a box, imply a discrete spectrum with
momentum $p=n/R$ and mass squared $m^2=n^2/R^2$. In any case there are the towers of KK recurrences of the graviton. They
are gravitationally coupled but there are a lot of them that sizably couple, so that the net result is a modification
of cross-sections and the presence of missing energy. 
There are many versions of these models. The SM brane can itself have a
thickness $r$ with $r<\sim 10^{-17}{\rm cm}$ or $1/r>\sim 1{\rm TeV}$, because we know that quarks and leptons are pointlike down to
these distances, while for gravity in the bulk there is no experimental counter-evidence down to $R<\sim 0.1 {\rm mm}$ or
$1/R>\sim 10^{-3}~eV$. In case of a thickness for the SM brane there would be KK recurrences for SM fields, like $W_n$,
$Z_n$ and so on in the TeV region and above. Large extra dimensions provide an exciting scenario. Already it is remarkable that this possibility is
compatible with experiment. However, there are a number of criticisms that can be brought up. First, the hierarchy problem is
actually translated in a new form rather than solved. In fact the basic relation $Rm=(M_{Pl}/m)^{2/n}$ shows that $Rm$, which one
would apriori expect to be $0(1)$, is instead ad hoc related to the large ratio $M_{Pl}/m$. Also it is
not clear how extra dimensions can by themselves solve the LEP paradox (the large top loop corrections should be
controlled by the opening of the new dimensions and the onset of gravity): since
$m_H$ is light
$\Lambda\sim 1/R$ must be relatively close. But precision tests put very strong limits on $\Lambda$. In fact in typical
models of this class there is no mechanism to sufficiently quench the corrections.

More recently models based on the Randall-Sundrum (RS) solution for the metric have attracted most of the model builders attention \cite{RS,FeAa}.  In these models the metric is not factorized and an exponential "warp" factor multiplies the ordinary 4-dimensional coordinates in the metric:
$ds^2=e^{-2kR\phi} \eta_{\mu \nu}dx^{\mu}dx^{\nu}-R^2\phi^2$  where $\phi$ is the extra coordinate. This non-factorizable metric is a solution of Einstein equations with specified 5-dimensional cosmological term. Two 4-dimensional branes are often localized at $\phi=0$ (the Planck or ultraviolet brane) and at $\phi=\pi$ (the infrared brane). In the simplest models all SM fields are located on the infrared brane. All 4-dim masses $m_4$ are scaled down with respect to
5-dimensional masses $m_5 \sim k \sim M_{Pl}$ by the warp factor: $m_4=M_{Pl}e^{-kR\pi}$. In other words mass and energy on the infrared brane are redshifted by the $\sqrt{g_{00}}$ factor. The hierarchy suppression $m_W/M_{Pl}$ could arise from the warping exponential $e^{-kR\phi}$, for not too large values of the warp factor exponent: $kR\sim 12$ (extra dimension are not "large" in this case). The question of whether these values of $kR$ can be stabilized has been discussed in ref.\cite{GW}. It was shown that the determination of $kR$ at a compatible value can be assured by a scalar field in the bulk ("radion") with a suitable potential which offer the best support to the solution of the hierarchy problem in this context. In the original RS models where the SM fields are on the brane and gravity is in the bulk there is a tower of spin-2 KK graviton resonances. Their couplings to ordinary particles are of EW order (because their propagator masses are red shifted on the infrared brane) and universal for all particles. These resonances could be visible at the LHC. Their signature is spin-2 angular distributions and universality of couplings. The RS original formulation is very elegant but 
when going to a realistic formulation it has problems, for example with EW precision tests. Also, in a description of physics from $m_W$ to $M_{Pl}$ there should be
place for GUTÕs. But, if all SM particles are on the {\rm TeV} brane the effective theory cut-off is low and no way to $M_{GUT}$ is open. Inspired by RS different realizations of warped geometry were tried: gauge fields in the bulk and/or all SM fields (except the Higgs) on the bulk. The hierarchy of fermion masses can be seen as the result of the different profiles of the corresponding distributions in the bulk: the heaviest fermions are those closest to the brane where the Higgs is located \cite{hoso}. 
While no simple, realistic model has yet emerged as a benchmark, it is attractive to imagine that ED could be a part of the truth, perhaps coupled with some additional symmetry or even SUSY.

Extra dimensions offer new possibilities for SUSY breaking. In fact, ED can realize a geometric separation between the hidden (on the Planck brane) and the visible sector (on the TeV brane), with gravity mediation in the bulk. In anomaly mediated SUSY breaking \cite{ano} 5-dim quantum gravity effects act as messengers. The name comes because
$L_{soft}$ can be understood in terms of the anomalous violation of a local superconformal invariance. In a particular formulation of 5 dimensional supergravity, at the 
classical level, the soft term are exponentially suppressed on the MSSM
brane. SUSY breaking effects only arise at quantum level through beta
functions and anomalous dimensions of the brane couplings and fields. In this case gaugino masses are proportional to gauge coupling beta functions, so that the gluino is much heavier than the electroweak gauginos. 

In the  general context of extra dimensions an interesting direction of development is the study of symmetry breaking by orbifolding and/or boundary conditions. Orbifolding means that we have a 5 (or more) dimensional theory where the extra dimension $x_5=y$ is compactified. Along $y$ one or more $Z_2$ reflections are defined, for example $P= y \leftrightarrow -y$ 
(a reflection around the horizontal diameter) and $P'= y \leftrightarrow -y-\pi R$ (a reflection around the vertical diameter). A field $\phi(x_\mu, y)$ with definite $P$ and $P'$ parities can be Fourier expanded along $y$. Then $\phi_{++}, \phi_{+-}, \phi_{-+}, \phi_{--}$ have the n-th Fourier components proportional to $\cos{\frac{2ny}{R}}, \cos{\frac{(2n+1)y}{R}}, \sin{\frac{(2n+1)y}{R}}, \sin{\frac{(2n+2)y}{R}} $, respectively. On the branes located at the fixed points of $P$ and $P'$, $y=0$ and $y= -\pi R/2$, the symmetry is reduced: indeed at $y=0$ only $\phi_{++}$ and $\phi_{+-}$ are non vanishing and only $\phi_{++}$ is massless. 

For example, at the GUT scale, symmetry breaking by orbifolding can be applied to obtain a reformulation of SUSY GUT's where many problematic features of ordinary GUT's (e.g. a baroque Higgs sector, the doublet-triplet splitting problem, fast proton decay etc) are eliminated or improved\cite{Kaw},\cite{edgut}. In these GUT models the metric is factorized, but while for the hierarchy problem $R\sim 1/{\rm TeV}$, here one considers $R\sim 1/M_{GUT}$ (not so large!). $P$ breaks $N=2$ SUSY, valid in 5 dimensions, down to $N=1$ while $P'$ breaks SU(5). At the weak scale there are models where SUSY, valid in $n>4$ dimensions, is broken by orbifolding \cite{ant}, in particular the model of ref.\cite{bar}, where the mass of the Higgs is in principle computable and is predicted to be light.

Symmetry breaking by boundary conditions (BC) is more general than the particular case of orbifolding\cite{groj}. Breaking by orbifolding is somewhat rigid: for example, normally the rank remains fixed and it corresponds to Higgs bosons in the adjoint representation (the role of the Higgs is taken by the 5th component of a gauge boson). BC allow a more general breaking pattern and, in particular, can lower the rank of the group. In a simplest version one starts from a 5 dimensional model with two branes at $y=0,~\pi R$. In the action there are terms localised on the  branes that also should be considered in the minimization procedure. For a scalar field $\varphi$ with a mass  term ($M$) on the boundary, one obtains the  Neumann BC $\partial_y \varphi=0$ for $M\rightarrow 0$ and the Dirichlet BC $\varphi=0$ for $M\rightarrow \infty$. In gauge theories one can introduce Higgs fields on the brane that take a VEV. The crucial property is that the gauge fields take a mass as a consequence of the Higgs mechanism on the boundary but the mass remains finite when the Higgs VEV goes to infinity. Thus the Higgs on the boundary only enters as a way to describe and construct the breaking but actually can be removed and still the gauge bosons associated to the broken generators take a finite mass. One is then led to try to formulate "Higgsless models" for EW symmetry breaking based on BC \cite{Hless} (some alternatives in 4 dimensions have also been described \cite{dec}which extend the old BESS model \cite{bess}). The RS warped geometry can be adopted with the Planck and the infrared branes. There is a larger gauge symmetry in the bulk which is broken down to different subgroups on the two branes so that finally of the EW symmetry only $U(1)_Q$ remains unbroken. The $W$ and $Z$ take a mass proportional to $1/R$. Dirac fermions are on the bulk and only one chirality has a zero mode on the SM brane. In Higgsless models unitarity, which in general is violated in the absence of a Higgs, is restored by exchange of infinite KK recurrences, or the breaking is delayed by a finite number, with cancellations guaranteed by sum rules implied by the 5-dim symmetry. Actually no compelling, realistic Higgsless model for EW symmetry breaking emerged so far. There are serious problems from EW precision tests \cite{BPR} because the smallness of the $W$ and $Z$ masses forces $R$ to be rather small and, as a consequence, the spectrum of KK recurrences is quite close. However these models are interesting as rare examples where no Higgs would be found at the LHC but instead new signals appear (new vector bosons, i.e. KK recurrences of the $W$ and $Z$).

An interesting model that combines the idea of the Higgs as a Goldstone boson and warped extra dimensions was proposed and studied in references \cite{con} with a sort of composite Higgs in a 5-dim AdS theory. It can be considered as a new way to look at walking technicolor \cite{L-C} using AdS/CFT correspondence. In a RS warped metric framework all SM fields are in the bulk but the Higgs is localised near the TeV brane. The Higgs is a pseudo-Goldstone boson (as in Little Higgs models) and EW symmetry breaking is triggered by 
top-loop effects. In 4-dim the bulk appears as a strong sector.  The 5-dimensional theory is weakly coupled so that the Higgs potential and EW observables can be computed.
The Higgs is rather light: $m_H < 185~{\rm GeV}$. Problems with EW precision tests and the $Zb \bar b$ vertex have been fixed in latest versions. The signals at the LHC for this model are 
a light Higgs and new resonances at ~1- 2 TeV

In conclusion, note that apart from Higgsless models (if any?) all theories discussed 
here have a Higgs in LHC range (most of them light).

\section{Effective theories for compositeness}  

In this approach \cite{comp}, \cite{comp2} a low energy theory from truncation of some UV completion is described in terms of an elementary sector (the SM particles minus the Higgs), a composite sector (including the Higgs, massive vector bosons $\rho_\mu$ and new fermions) and a mixing sector. The Higgs is a pseudo Goldstone boson of a larger broken gauge group, with $\rho_\mu$ the corresponding massive vector bosons. Mass eigenstates are mixtures of elementary and composite states, with light particles mostly elementary and heavy particles mostly composite. But the Higgs is totally composite (perhaps also the right-handed top quark: in general the top sector can reveal observable signatures of compositeness at the LHC \cite{comp3}). New physics in the composite sector is well hidden because light particles have small mixing angles. The Higgs is light because only acquires
mass through interactions with the light particles from their composite components. This general description can apply to models with a strongly interacting sector as arising from little Higgs or extra dimension scenarios.

\section{The anthropic solution}

The apparent value of the cosmological constant $\Lambda$ poses a tremendous, unsolved naturalness problem \cite{tu}. Yet the value of $\Lambda$ is close to the Weinberg upper bound for galaxy formation \cite{We}. Possibly our Universe is just one of infinitely many (Multiverse) continuously created from the vacuum by quantum fluctuations. Different physics takes place in different Universes according to the multitude of string theory solutions (~$10^{500}$). Perhaps we live in a very unlikely Universe but the only one that allows our existence \cite{anto},\cite{giu}. I find applying the anthropic principle to the SM hierarchy problem excessive. After all we can find plenty of models that easily reduce the fine tuning from $10^{14}$ to $10^2$: why make our Universe so terribly unlikely? By comparison the case of the cosmological constant is a lot different: the context is not as fully specified as the for the SM (quantum gravity, string cosmology, branes in extra dimensions, wormholes through different Universes....)

\section{Conclusion}

Supersymmetry remains the standard way beyond the SM. What is unique to SUSY, beyond leading to a set of consistent and
completely formulated models, as, for example, the MSSM, is that this theory can potentially work up to the GUT energy scale.
In this respect it is the most ambitious model because it describes a computable framework that could be valid all the way
up to the vicinity of the Planck mass. The SUSY models are perfectly compatible with GUT's and are actually quantitatively
supported by coupling unification and also by what we have recently learnt on neutrino masses. All other main ideas for going
beyond the SM do not share this synthesis with GUT's. The SUSY way is testable, for example at the LHC, and the issue
of its validity will be decided by experiment. It is true that we could have expected the first signals of SUSY already at
LEP, based on naturality arguments applied to the most minimal models (for example, those with gaugino universality at
asymptotic scales). The absence of signals has stimulated the development of new ideas like those of extra dimensions
and "little Higgs" models. These ideas are very interesting and provide an important reference for the preparation of LHC
experiments. Models along these new ideas are not so completely formulated and studied as for SUSY and no well defined and
realistic baseline has sofar emerged. But it is well possible that they might represent at least a part of the truth and it
is very important to continue the exploration of new ways beyond the SM. New input from experiment is badly needed, so we all look forward to the start of the LHC.

The future of particle physics heavily depends on the outcome of the LHC. So the questions that many people ask are listed in the following with my (tentative) answers. Is it possible that the LHC does not find the Higgs particle? Yes, it is possible, but then must find something else (experimental and theoretical upper bounds on the Higgs mass in the SM, unitarity violations in the absence of a Higgs or of new physics). Is it possible that the LHC finds the Higgs particle but no other new physics (pure and simple SM)? Yes, it is  technically possible but it is not natural (would go in the direction that we live in a very eccentric Universe). Is it possible that the LHC finds neither the Higgs nor new physics? No, it is Òapproximately impossibleÓ (meaning that the only possible way out would be that the LHC energy is a bit too low and only misses by a small gap the onset of the solution).

\begin{acknowledgement}

I am very grateful to George Zoupanos and the Organising Committee for inviting me at the Corfu Institute 2009 and for their kind hospitality. This research was financed in part by the Marie Curie Research Training Network MRTN-CT-2006-035505, "Tools and Precision Calculations for Physics Discoveries at Colliders".

\end{acknowledgement}


\begin{thebibliography}{[1]}

\bibitem{sprin} An updated review of the SM can be found in G.~Altarelli, in  Elementary Particles (Landolt-Bšrnstein, New Series / Elementary Particles, Nuclei and Atoms), ed. by H. Schopper, Springer, Berlin, 2008, chap.~2, 3 and 4, 
http://www.fis.uniroma3.it/it.php?page=Didattica\&argo=Laurea\%20Magistrale\&cat=Corsi\&id=661\&anno=
\bibitem{djou1} A.~Djouadi, Phys. Rept. \textbf{457} 1  (2008), Archiv:hep-ph/0503172.
\bibitem{wells} For a recent pedagogical review, see J.D.~Wells, ArXiv:0909.4541.
\bibitem{ref:unit} B. W.~Lee, C.~Quigg and H.B.~Thacker  {Phys.Rev.} {\textbf D16} {1519}  (1977).
\bibitem{ref:isid} {G.~Isidori, G.~Ridolfi and A.~Strumia~}  {Nucl. Phys.} {\textbf B609}{387} (2001) and references therein.
\bibitem{ref:hri} {T.~Hambye and K.~Riesselmann}  {Phys.Rev.}{\textbf D55} {7255} (1997)  and references therein.
\bibitem{ref:ae1}  {B.~Odom~{\it et~al.}} {Phys. Rev. Lett.} {\textbf 97} {030801} (2006).
\bibitem{ref:amu} {Muon g-2 Collab., G.W.~Bennett~{\it et~al.}} {Phys. Rev.}{\textbf D73}{072003} (2006) .
\bibitem{ref:ae2} {G.~Gabrielse~et~al} {Phys.Rev.Lett.} {\textbf 97}
{030802} (2006); D.~Hanneke, S.~Fogwell, and G.~Gabrielse, Phys. Rev. Lett. {\textbf 100} 120801 (2008), ArXiv:0801.1134. 
\bibitem{ref:kino} T.~Aoyama {\it et~al.}, ArXiv:1001.3704 and and references therein.
\bibitem{ref:amu2} {M.~Passera}, ArXiv: hep-ph/0702027; {F.~Jegerlehner}, ArXiv: hep-ph/0703125; {D.W.~Hertzog~{\it et~al.}}, ArXiv:0705.4617; K.~Hagiwara {\it et~al.}, Phys. Lett. {\textbf B649} 173 (2007), ArXiv:hep-ph/0611102; M.~Davier {\it et~al.}, ArXiv:0906.5443;
M.~Davier {\it et~al.}, ArXiv:0908.4300. 
\bibitem{ref:ewg} The LEP Electroweak Group, http://lepewwg.web.cern.ch/LEPEWWG/.
\bibitem{ref:AG} {G.~Altarelli and M.~Grunewald} {Phys.Rept.} {\textbf 189}{403} (2004), ArXiv: hep-ph/0404165.
\bibitem{SJindariani} S.~Jindariani, Proceedings of the Hadron Collider Physics Symposium, Evian, France, November 2009.
\bibitem{rept} For a recent review, see M.~Antonelli {\it et~al.}, ArXiv:0907.5386.
\bibitem{isid} For a recent review, see  G.~Isidori, ArXiv: 1001.3431.
\bibitem{revnu} For reviews, see, for example, G.~Altarelli and F.~Feruglio,
New J.\ Phys.\  {\textbf 6} (2004) 106, ArXiv:hep-ph/0405048; G.~Altarelli, ArXiv:0905.3265;
R.N.~Mohapatra and A.Y.~Smirnov,
Ann. Rev. Nucl. Part. Sci. {\textbf 56} 569 (2006)
ArXiv:hep-ph/0603118;
W.~Grimus, PoS P2GC:001,2006. 
ArXiv:hep-ph/0612311;
R.N.~Mohapatra {\it et~al.}, Rept.Prog.Phys. {\textbf 70}1757 (2007), ArXiv:hep-ph/0510213;
M.C.~Gonzalez-Garcia and M.~Maltoni, Phys. Rept.  {\textbf 460}  1 (2008), ArXiv:0704.1800.
\bibitem{ref:buch} For a general review, see, for example, {W.~Buchmuller, R.D.~Peccei and T.~Yanagida}, {Ann.Rev.Nucl.Part.Sci.}{\textbf 55}{311} (2005), ArXiv:hep-ph/0502169.
\bibitem{datanu} G.L.~Fogli {\it et~al.}, Phys.\ Rev.\ Lett.\  {\textbf 101} (2008) 141801
ArXiv:0806.2649; G.L.~Fogli {\it et~al.},
Nucl.Phys.Proc.Suppl.188:27,2009,
ArXiv:0809.2936;
T.~Schwetz, M.~Tortola and J.W.~Valle,
New J. Phys.  {\textbf 10} (2008) 113011
ArXiv:0808.2016;
M.~Maltoni and T.~Schwetz,  ArXiv:0812.3161.
\bibitem{Harr} P.F.~Harrison, D.H.~Perkins and W.G.~Scott,
Phys. Lett. B {\textbf 530}167  (2002),
ArXiv:hep-ph/0202074; P.F.~Harrison and W.G.~Scott,
Phys. Lett. B {\textbf 535} 163  (2002),
ArXiv:hep-ph/0203209; 
Phys. Lett. B {\textbf 557}  76 (2003),
ArXiv:hep-ph/0302025; ArXiv:hep-ph/0402006.
\bibitem{rmp} For a review, see G.~Altarelli and F.~Feruglio, ArXiv:1002.0211.
\bibitem{MEG} The MEG collaboration: J. Adam  {\it et~al.}, ArXiv:0908.2594
\bibitem{oli} For a recent summary, see K. A.~Olive, ArXiv:1001.5014.
\bibitem{ber} R.~Bernabei {\it et~al.} [DAMA Collab.], Eur. Phys. J. {\textbf 205} 333 (2008). 
\bibitem{pam} O.~Adriani {\it et~al.} [PAMELA Collab.], Nature {\textbf 458} 607 (2009), ArXiv: 0810.4995. 
\bibitem{atic} J.~Chang {\it et~al.} [ATIC Collab.], Nature {\textbf 456}  362 (2008). 
\bibitem{fer} A.A.~Abdo {\it et~al.} [Fermi-LAT Collab.], Phys. Rev. Lett. {\textbf 102} 181101  (2009), ArXiv:0905.0025.
\bibitem{exdm} V.~Barger {\it et~al.}, ArXiv: 0809.0162; M.~Cirelli {\it et~al.}, ArXiv:0802.3378; 0809.2409; N.~Arkani-Hamed {\it et~al.}, ArXiv:0810.0713; JHEP {\textbf 0812} 104 (2008), ArXiv:0810.0714, E.~Nardi {\it et~al.}, 
ArXiv:0811.4153, G.~Bertone {\it et~al.}, ArXiv:0811.3744; E.~Ponton and L.~Randall, ArXiv:0811.1029; P.~Meade {\it et~al.}, ArXiv:0905.0480; M.~Cirelli, P.~Panci and P.\,D. Serpico, ArXiv:0912.0663 and references therein. 
\bibitem{astr} See, for example, I.~Busching {\it et~al.}, Astrophys. J. {\textbf 78} L39 (2008). 
\bibitem{hie} G.~Õt Hooft, NATO Adv. Study Inst. Ser. B Phys. {\textbf 59} 135 (1980) ; M.\,J. G. Veltman, Acta Phys. Polon. {\textbf B 12}  437 (1981). 
\bibitem{giu} G.~Giudice,  in "Perspectives on LHC Physics", ed. by G.~Kane and A.~Pierce, World Sci., ArXiv:0801.2562.
\bibitem{ref:BS} {R.~Barbieri and A.~Strumia}, ArXiv:hep-ph/0007265.
\bibitem{CGrojean} For a recent overview see C.~Grojean, ArXiv:0910.4976. 
\bibitem{Martin} For recent pedagogical reviews see, for example, S.\,P. Martin, hep-ph/9709356; I.J. R. Aitchison, ArXiv:hep-ph/0505105; M.~Drees, R.~Godbole and P.~Roy, {\it Theory and Phenomenology of Sparticles},  World Sci. (2005). 
\bibitem{mas} For a recent review of flavour and GUT's, see, for example, A.~Masiero, S.\,K. Vempati and O. ~Vives, ArXiv:0711.2903.
\bibitem{extsusy} T.~Gherghetta, C.\,F. Kolda and S.\,P. Martin, Nucl. Phys.  {\textbf B 468}  37 (1996), ArXiv:hep-ph/9510370; A. Strumia, Phys. Lett.  {\textbf B 466}  107 (1999), ArXiv:hep-ph/9906266; A.~Brignole {\it et~al.}, Nucl. Phys.  {\textbf B 666}  105 (2003), ArXiv:hep-ph/0301121; P.~Batra {\it et~al.}, JHEP  {\textbf 0402}  043 (2004) ArXiv:hep-ph/0309149; A.~Birkedal, Z. ~Chacko and M. K. Gaillard, JHEP {\textbf 0410}  036 (2004), ArXiv:hep-ph/0404197; P.\'H. Chankowski {\it et~al.}, Phys. Lett. {\textbf B 598}  252 (2004), ArXiv:hep-ph/0407242; M.~Dine, N. ~Seiberg and S.~Thomas, Phys. Rev. {\textbf D 76}  095004 (2007), ArXiv:0707.0005;  I.~Antoniadis, E.~Dudas and D.\,M. Ghilencea, JHEP {\textbf 0803}  045 (2008), ArXiv:0708.0383; K.~Blum and Y.~Nir, Phys. Rev. {\textbf D 78} 035005 (2008), ArXiv:0805.0097; I.~Antoniadis {\it et~al.}, Nucl. Phys. {\textbf B 808} 155 (2009), ArXiv:0806.3778;   K.~Blum, C.~Delaunay and Y.~Hochberg, ArXiv:0905.1701;  N. Bernal {\it et~al.}, JHEP {\textbf 0908}  053 (2009), ArXiv:0906.4696; M. Carena {\it et~al.}, ArXiv:0909.5434. 
\bibitem{nmssm} H.\,P. Nilles, M.~Srednicki and D.~Wyler, Phys. Lett.  \textbf{B120}, 346 (1983);
J.\,P. Derendinger and C.\,A. Savoy, Nucl.  Phys.  \textbf{B 237}, 307 (1984);
M.~Drees, Int. J. Mod. Phys. \textbf{A4}, 3635 (1989);
J.~R. Ellis et al, Phys. Rev. \textbf{ D39},  844 (1989);
T.~Elliott, S.\,F. King and P.\,L. White, Phys. Lett. \textbf{B314}, 56 (1993),
ArXiv:hep-ph/9305282; Phys.  Rev.  \textbf{D49}, 2435 (1994),
ArXiv:hep-ph/9308309; U.~Ellwanger, M.~Rausch de Traubenberg and C.\, A. Savoy,
Phys. Lett.  \textbf{B315}, 331 (1993), ArXiv:hep-ph/9307322;
B.\,R. Kim, A.~Stephan and S.\,K. Oh, Phys. Lett.  \textbf{B336}  200 (1994).
\bibitem{barbie} R.~Barbieri {\it et~al.} Phys.Rev. {\textbf D75} 035007 (2007), ArXiv:hep-ph/0607332.
\bibitem{gau} M.~Dine and A.\,E. Nelson, Phys. Rev. \textbf{D48},  1277 (1993);
M.~Dine, A.~E. Nelson and Y.~Shirman, Phys. Rev. \textbf{D51}, 1362 (1995);
G.\,F. Giudice and R.~Rattazzi, Phys. Rept. \textbf{322}, 419 (1999).
\bibitem{LP} P.~Langacker and N.~Polonsky, Phys. Rev. \textbf{D52}, 3081 (1995). See also D.~ Ghilencea, M.~ Lanzagorta and G.G.~Ross, Nucl.Phys. \textbf{B511}, 3 (1998).
ArXiv:hep-ph/9707401; D.~Ghilencea, G.G.~Ross, Nucl.Phys. \textbf{B606},101 (2001), ArXiv: hep-ph/0102306.
\bibitem{pdg} W.-M. Yao et al., The Review of Particle Physics,
Journ. of Phys. \textbf{G 33}, 1 (2006).
\bibitem{AFM} G.~Altarelli, F.~Feruglio and I.~Masina, JHEP \textbf{0011}, 040 (2000).
\bibitem{sus} O.~Buchmueller et al, Phys.Lett. \textbf{B657}, 87(2007), ArXiv:0707.3447;
J.\,R. Ellis et al, JHEP, {0708} 083 (2007), ArXiv:0706.0652.
\bibitem{djou} A.~ Djouadi, Phys. Rept. \textbf{459} 1 (2008), ArXiv:hep-ph/0503173.
\bibitem{lephwg} The LEP Higgs Working Group,
http://lephiggs.web.cern.ch/LEPHIGGS/www/Welcome.html
\bibitem{ross} Fpr a recent analysis see S.~Cassel, D.\,M. Ghilencea and  G.\,G. Ross, ArXiv:0911.1134; ArXiv:1001.3884 and references therein.
\bibitem{kane} G.~Kane et al, Phys. Lett. \textbf{ B551},146 (2003).
\bibitem{gium} G.\,F. Giudice and A.~Masiero, Phys. Lett. \textbf{B206}, 480 (1988).
\bibitem{ref:L-C} For reviews see, for example, {K.~Lane}, ArXiv:hep-ph/0202255; {R.S.~Chivukula}, ArXiv:hep-ph/0011264.
\bibitem{schm} N. ~Arkani-Hamed, A.\,G. Cohen and H.~Georgi, Phys. Lett. \textbf{B 513}  232 (2001), ArXiv:hep-ph/0105239. For reviews and a list of references, see, for example, M.~Schmaltz and D.~Tucker-Smith, ArXiv:hep-ph/0502182; M. Perelstein, ArXiv:hep-ph/0703138. 
\bibitem{Ch} H-C.~Cheng and I.~Low, JHEP  \textbf{0408} (2004) 061, ArXiv:hep-ph/0405243; J.~Hubisz et al, JHEP \textbf{0601} (2006) 135, ArXiv:hep-ph/0506042. 
\bibitem{hill} C.\,T. Hill and R.\,J. Hill,   Phys.Rev. \textbf{D76},115014 (2007), 
ArXiv:0705.0697; C.~Csaki, J.~Heinonen, M.~Perelstein and C. ~Spethmann, ArXiv:0804.0622.
\bibitem{sign} M.\,S. Carena {\it et~al.}, Phys. Rev. \textbf{D 75}  091701 (2007), ArXiv:hep-ph/0610156; T. ~Han {\it et~al.}, Phys. Rev. \textbf{D 67}  095004 (2003), ArXiv:hep-ph/0301040; M.~Perelstein, M.\,E. Peskin and A.~Pierce, Phys. Rev. textbf{D 69}  075002 (2004), ArXiv:hep-ph/0310039. 
\bibitem{led} N.~Arkani-Hamed, S.~Dimopoulos and G.\,R. Dvali,  Phys. Lett. \textbf{B 429}, 263 (1998), ArXiv:hep-ph/9803315;  Phys. Rev. \textbf{D 59}, 086004 (1999); ArXiv:hep-ph/9807344; 
I. Antoniadis {\it et~al.}, Phys. Lett. \textbf{B436}, 257 (1998), ArXiv:hep- ph/9804398. 
\bibitem{Jo} For a review and a list of references, see, for example, J.~Hewett and M.~Spiropulu,  Ann. Rev. Nucl. Part. Sci. \textbf{52}, 397 (2002), ArXiv:hep-ph/0205196. 
\bibitem{cav} C.\,D. Hoyle {\it et~al.}, Phys. Rev. Lett. \textbf{86}, 1418 (2001), ArXiv:hep-ph/0011014; E. G. Adelberger et al [EOT-WASH Group Collaboration], ArXiv:hep-ex/0202008. 
\bibitem{RS} L.~Randall and R.~Sundrum, Phys. Rev. Lett. \textbf{ 83}, 3370 (1999), \textbf{83}, 4690 (1999).
\bibitem{FeAa} For pedagogical reviews, see for example, R.~Sundrum, ArXiv:hep-th/0508134; R.~Rattazzi, ArXiv:hep-ph/0607055; C.~Csaki, J.~Hubisz and P.~Meade, ArXiv:hep-ph/0510275; T.~Gherghetta, ArXiv:hep-ph/0601213.
\bibitem{GW} W.\,D. Goldberger and M.\,B. Wise, Phys. Rev. Letters \textbf{83}, 4922 (1999), ArXiv:hep-ph/9907447.
\bibitem{ano} L.~Randall and R.~Sundrum, Nucl. Phys. \textbf{B557}, 79 (1999); G.\,F. Giudice et al, JHEP \textbf{9812}, 027 (1998).
\bibitem{ant} I. ~Antoniadis, C.~Munoz and M.~Quiros, Nucl.Phys. \textbf{B397},  515 (1993);
A.~Pomarol and M.~Quiros,  Phys. Lett. \textbf{B438},  255 (1998).
\bibitem{hoso}  See for example,  C.\,A. Scrucca, M.\,Serone and L.\,Silvestrini, Nucl.Phys. {\textbf B669}, 128 (2003), ArXiv:hep-ph/0304220 and references therein. See also Y.~Grossman and M.~Neubert, Phys.Lett. \textbf{B474}, 361 (2000), ArXiv:hep-ph/9912408; T.~Gherghetta and A.~Pomarol, Nucl.Phys. \textbf{B586},141 (2000), ArXiv:hep-ph/0003129.
\bibitem{bar} R.~Barbieri, L.~Hall and Y.~Nomura, Nucl.Phys. \textbf{B624}, 63 (2002); R.~Barbieri, G. ~Marandella and M.~Papucci, Phys.Rev. {\textbf D66}, 095003 (2002), ArXiv:hep-ph/0205280; Nucl.Phys. {\textbf B668} 273 (2003),  ArXiv:hep-ph/0305044  and references therein. See, however, D.~Ghilencea, JHEP {\textbf 0503}, 009 (2005), ArXiv: hep-ph/0409214, D.~Ghilencea, H.M.~Lee, JHEP {\textbf 0509}, 024 (2005), ArXiv: hep-ph/0505187. 
\bibitem{groj} For a pedagogical introduction, see C.~Grojean, CERN-PH-TH/2006-172.
\bibitem{Hless} See for example, C.~Csaki {\it et~al.}, Phys.Rev. {\textbf D69}, 055006 (2004), ArXiv:hep-ph/0305237; Phys.Rev.Lett. {\textbf 92}, 101802 (2004), ArXiv:hep-ph/0308038; Phys.Rev. {\textbf D70}, 015012 (2004), ArXiv:hep-ph/0310355; S.~Gabriel, S.~Nandi and G.~Seidl, {\it Phys.Lett.} {\textbf B603}, 74 (2004), ArXiv:hep-ph/0406020 ; R.~Chivukula {\it et~al.}, Phys.Rev. {\textbf D74}, 075011 (2006), ArXiv:hep-ph/0607124;  Phys.Rev. {\textbf D78}, 077701 (2008), ArXiv:0808.2071 and references therein.
\bibitem{dec} A.\.S. Belyaev {\it et~al.}, ArXiv:0907.2662; E. Accomando {\it et~al.}, Nuovo Cim. 123B 809 (2008), ArXiv:0807.2951. 
\bibitem{bess} R. Casalbuoni {\it et~al.}, Phys. Lett. B 155 (1985) 95
\bibitem{BPR} R.~Barbieri, A.~Pomarol and R.~Rattazzi,  Phys.Lett. {\textbf B591}, 141 (2004), ArXiv:hep-ph/0310285;  G.~Cacciapaglia {\it et~al.}, Phys. Rev. {\textbf D 70}  075014 (2004),ArXiv:hep-ph/0401160. 
\bibitem{Kaw} E.~Witten, Nucl. Phys.  {\textbf B258}  75 (1985); Y.~Kawamura, Progr. Theor. Phys. \textbf{105}, 999 (2001); A.\,E. Faraggi, Phys. Lett.   {\textbf B520}  337 (2001), ArXiv:hep-ph/0107094.
\bibitem{edgut} G.~Altarelli and F.~Feruglio, Phys.Lett. {\textbf B511} 257 (2001),
ArXiv:hep-ph/0102301; L.\,J. Hall and Y.~Nomura, Phys. Rev.  {\textbf  D64} (2001) 055003, ArXiv:hep-ph/0103125; Phys.Rev.  {\textbf  D66}  075004 (2002), ArXiv:hep-ph/0205067; A.~Hebecker and J.~March-Russell, Nucl. Phys. {\textbf  B613} (2001) 3, ArXiv:hep-ph/0106166; Phys. Lett.  {\textbf  B541}  338 (2002), ArXiv:hep-ph/0205143;
T.~Asaka, W.~Buchmuller and  L.~Covi, Phys.Lett.  {\textbf B523},199 (2001), ArXiv:hep-ph/0108021.
\bibitem{prob} J.\L. Hewett, F.\,J. Petriello and T.\,G. Rizzo, JHEP {\textbf 0310} 062 (2003), ArXiv:hep-ph/0211218; C.~Csaki {\it et~al.}, Phys.Rev. {\textbf D67} 115002 (2003), ArXiv:hep-ph/0211124; Phys.Rev. {\textbf D68}, 035009 (2003), ArXiv:hep-ph/0303236.
\bibitem{con} R.~Contino, Y.~Nomura, and A.~Pomarol, Nucl. Phys. \textbf{B671}, 148 (2003), Arxiv:hep-ph/0306259; K.~Agashe, R.~Contino and A.~Pomarol, Nucl.Phys. \textbf{B719}165 (2005), ArXiv:hep-ph/0412089; K.~Agashe, R.~Contino, Nucl.Phys. \textbf{B742} 59 (2006), ArXiv:hep-ph/0510164; K. Agashe {\it et~al.},  Phys.Lett. {\textbf B641} 62 (2006), ArXiv:hep-ph/0605341.
\bibitem{L-C} K.~Lane, ArXiv:hep-ph/0202255; R.\,S. Chivukula, ArXiv:hep-ph/0011264.
\bibitem{comp} R.~Contino {\it et~al.}, JHEP \textbf{0705},074 (2007), ArXiv:hep-ph/0612180; G.F. Giudice et al {\it JHEP} \textbf{0706}, 045 (2007), ArXiv:hep-ph/0703164.
\bibitem{comp2} R.~Barbieri {\it et~al.}, Phys. Rev. \textbf{D 76}  115008 (2007),  ArXiv:0706.0432 [hep-ph; 
C.~Anastasiou, E.~Furlan and J.~Santiago, ArXiv:0901.2117; 
R.~Contino, ArXiv:0908.3578; 
K.~Agashe and R.~Contino, Nucl. Phys. \textbf{B 742}  59 (2006), ArXiv:hep-ph/0510164; 
M.\,S. Carena {\it et~al.}, Phys. Rev. \textbf{D 76}  035006 (2007), ArXiv:hep-ph/0701055; 
M.~Gillioz, Phys. Rev. \textbf{D 80}  055003 (2009), ArXiv:0806.3450; 
C.~Bouchart and G.~Moreau, Nucl. Phys. \textbf{B 810}  66 (2009), ArXiv:0807.4461. 
R.~Contino, L.~Da Rold and A.~Pomarol, Phys. Rev. \textbf{D 75}  055014 (2007), ArXiv:hep-ph/0612048.
\bibitem{comp3} M.~Carena {\it et~al.}, Phys. Rev. \textbf{D 77} 
076003 (2008) , ArXiv:0712.0095; R.~Contino and G.~Servant, JHEP \textbf{0806} 026 (2008) , ArXiv:0801.1679; P. ~Lodone, JHEP \textbf{0812} 029 (2008) , ArXiv:0806.1472;  A.~Pomarol and J.~Serra, Phys. Rev. \textbf{D 78}  074026 (2008) , ArXiv:0806.3247; H.~de Sandes and R.~Rosenfeld, J. Phys. \textbf{G 36}  085001 (2009), ArXiv:0811.0984; J.~Mrazek and A.~Wulzer, ArXiv:0909.3977.
\bibitem{We} S.~Weinberg, Phys. Rev. Lett. \textbf{59},  2607 (1987). 
\bibitem{anto} N.~Arkani-Hamed and S.~Dimopoulos, {\it JHEP} {\textbf 0506}, 073 (2005), ArXiv:hep-th/0405159; N.~Arkani-Hamed et al, {\textbf Nucl.Phys.} {\textbf B709} 3 (2005), ArXiv:hep-ph/0409232; G. ~Giudice and A.~Romanino, {\it Nucl.Phys.} {\textbf B699}, 65 (2004), Erratum-ibid. {\textbf B706}, 65 (2005), ArXiv:hep-ph/0406088; N.~Arkani-Hamed, S.~Dimopoulos, S.~Kachru, ArXiv:hep-ph/0501082;
G.~Giudice, R.~Rattazzi, {\it Nucl.Phys.} {\textbf B757}, 19 (2006), ArXiv:hep-ph/0606105.



\end{thebibliography}
\end{document}